\documentclass[a4paper,fleqn,usenatbib]{mnras}

\usepackage{newtxtext,newtxmath}

\usepackage[T1]{fontenc}
\usepackage{ae,aecompl}

\usepackage{graphicx}	
\usepackage{amsmath}	
\usepackage{amssymb}	
\usepackage{array}

\title[HAT-P-19 System]{A Holistic and Probabilistic Approach to the Ground-based Data of HAT-P-19 System}

\author[O. Basturk et al.]{
\"Ozg\"ur Ba\c{s}t\"urk,$^{1}$\thanks{E-mail: obasturk@ankara.edu.tr}
S. Yal\c{c}{\i}nkaya,$^{1}$
E. M. Esmer,$^{1}$
T. Tanr{\i}verdi$^{2}$
and B. Keten$^{1}$
\\
$^{1}$Ankara University, Faculty of Science, Astronomy \& Space Sciences Department, Tandogan, TR-06100, Ankara, Turkey\\
$^{2}$ Ni\u{g}de \"{O}mer Halisdemir University, Faculty of Arts and Sciences, Department of Physics, TR-51240, Ni\u{g}de, Turkey
}

\date{Accepted XXX. Received YYY; in original form ZZZ}

\pubyear{2019}

\begin{document}
\label{firstpage}
\pagerange{\pageref{firstpage}--\pageref{lastpage}}
\maketitle

\begin{abstract}
We update the fundamental physical and orbital properties of the transiting hot-Saturn type exoplanet HAT-P-19b and its host star HAT-P-19 as a result of the global modeling of our high-precision transit light curves, an archive spectrum, radial velocity observations, brightness values from broadband photometry in different passbands, and the precise distance of the system derived from its Gaia parallax. We collected all the light curves obtained with ground-based photometry by amateur and professional observers, measured mid-transit times, analyzed their differences from calculated transit timings based on reference ephemeris information, which we update as a result. We haven't found any periodicity in the residuals of a linear trend, which we attribute to the accumulation of uncertainties in the reference mid-transit time and the orbital period. We discuss the potential origins of the variation in transit timings briefly and find stellar activity as the most likely cause. Finally, we comment on the scenarios describing the formation and migration of this hot-Saturn type exoplanet with a bloated atmosphere yet a small core, although it is orbiting a metal-rich ($[Fe / H] = 0.24$ dex) host star based on the planetary, orbital, and stellar parameters of the system that we derived from our global model, the age and the evolutionary history of the star. 
\end{abstract}




\section{Introduction}
\label{sec:introduction}
Transiting exoplanets are of interest for several reasons. First of all, many physical properties of the planet and the host star can be directly measured or derived based on a minimum set of assumptions and the established theory of stellar evolution. Furthermore, variations in their transit timings can be indicative of an unstable orbit or additional bodies gravitationally bound to the system. In addition, planet formation and migration mechanisms of especially the gas giants orbiting at short periods, so-called hot-Jupiters \& Saturns, can be tested based on a sample space containing well-characterized planets that have been discovered frequently in transit surveys due to the biases of the technique. As of now, even atmospheric constituents and properties of a handful of such planets have been constrained. But with the installment of new ground-based and space-borne telescopes, many more are in line for more detailed characterization in terms of their atmospheric properties as well. Therefore, there is great merit in observing these systems frequently, analyzing their archival data and studying them in more detail with the help of new and more advanced statistical tools and scientific approaches incorporated into recently developed computer code. From this perspective, we aim to present new and the most precise photometric observations of HAT-P-19 system from the ground so far, as well as to refine its parameters based on these observations and a renewed analysis of its high resolution spectra in the Keck / HIRES archive with a new spectroscopic analysis tool, and finally study the variations in the mid-transit times of the hot-Saturn type, transiting exoplanet HAT-P-19b by analyzing our own observations as well as the observations of professional and amateur observers. 

HAT-P-19b is a low-density ($\rho_p = 0.28$ g cm$^{-3}$), Saturn-mass ($M_p = 0.97$ M$_{Saturn}$) exoplanet orbiting a metal-rich ($[Fe / H] = +0.24$ dex) star, on a $P = 4.01$ day, small eccentricity ($e = 0.08$) orbit. Its host star is a typical K-type main-sequence star with $T_{eff} \sim 4960$ K, and $log~g = 4.57$. Confirmed transiting planets with a mass around that of Saturn have different densities, orbiting their host stars with diverse physical and orbital properties. Therefore, it is crucial to determine the parameters of such systems accurately and precisely to carry out a population analysis of the planets within this mass regime to understand their formation, orbital evolution, and the correlations between their parameters and that of their host stars. 

High iron abundance observed in the host star's atmosphere ($[Fe / H] \sim 0.24$ dex), in contrast of the low density of its planet ($\rho_p \sim 0.28 g~cm^{-3}$) makes it a contradicting example for the correlation assumed to be existing between the core mass of the planet and the metallicity of the star in this mass regime. Explaining such planets ($[Fe / H]_{\ast} > 0.20$, $\rho_{p} < 0.40$, $P_{orb} < 10$ days), number of which nears 10 at the moment, is a challenge for core-accretion scenario, requiring a migration mechanism to transport the planets from where they form to where they are observed now. However, a very recent study has asserted in-situ formation within core-accretion scenario as a potential mechanism for the formation of hot-Jupiters and Saturns \citep{bailey2018}.  While the observed radius anomaly in these hot \& bloated Saturns is reasonable considering the flux they are subject to from their host stars at their short orbital distances; core mass is still below the limits expected from the high metal content in such proximity of the super-solar metallicity stars \citep{felzting2001}, where they might be claimed to have formed. In the case of in-situ formation, kinetic heating \citep{guillot2002} might be the mechanism that can unravel the story behind the inflation of such planets orbiting metal-rich stars. HAT-P-19b is an important member of this class of planets whose cores are expected to be enriched in metal content, therefore, should have been larger if they form where they are observed. If they had formed outside the snow-line and have migrated inwards, then the reason behind the inflation of their atmosphere, and the type of their migration history will be the questions to answer. Consequently, determining the fundamental physical and orbital properties of the individual planets in this class and their host stars within great precision and accuracy is vitally important when these questions are addressed based on the parameter space from a sample of the individuals. 

HAT-P-19b's extended radius ($R_p \sim 1.09$ R$_{jup}$) for its mass ($M_p \sim 0.97$  M$_{saturn}$) also makes this hot \& inflated Saturn-mass planet an outstanding  target for transmission spectroscopy together with the recently found WASP-160Bb \citep{lendl2019}.  with similar orbital and physical characteristics, both orbiting metal-rich stars. So far, XO-2b is the only planet orbiting a metal-rich star ($[Fe / H] \sim 0.45$), \citep{teske2015}) with the detection of both Na and K in its atmosphere \citep{sing2011, pearson2019}. Another hot-Saturn type planet WASP-49b has also been reported to have aerosol constraints \citep{cubillos2017} and neutral sodium at its high-altitudes \citep{wyttenbach2017} of its extended atmosphere with cloud decks \citep{lendl2016}. Only \citet{mallonn2015} have attempted at obtaining a transmission spectrum of HAT-P-19 with the OSIRIS spectrograph at the Gran Telescopio Canarias back in 2012. They have not found any trace of an additional absorption at any wavelength or any slope in their differential spectrophotometric search. With the same instrument, \citet{sing2011} found a trace of potassium in the atmosphere of XO-2b. Nevertheless, HAT-P-19b is still a promising target in terms of its potential of bearing heavy elements such as sodium and potassium, which may have escaped detection because of the pressure broadening in the planetary atmosphere \citep{mallonn2015}. Whether such elements are found or not, this will provide another evidence for or against a suspected correlation between heavy element content in exoplanet atmospheres and the planetary mass \citep{nikolov2018}. These questions about the atmospheric content of this interesting planet, which is suggested as one of the prime targets to be observed with the James Webb Space Telescope (JWST) \citep{moliere2017}, can be answered with the transmission spectroscopy observations from the space in the future. Therefore, determination of the planet and host star properties as a result of a detailed analysis of existing and new data with a holistic approach making use of recently developed analysis codes and global modeling techniques will help in characterization of its atmosphere too. Refining its ephemeris information is also crucial for planning future observations of the target with JWST and ground-based telescopes.

The system is also of particular interest due to the linear trend observed in its radial velocity residuals \citep{hartman2011}, potentially indicating a gravitationally bound, yet unseen companion perturbing its orbit \citep{hartman2011, seeliger2015} and / or causing the arrival times of the light from the system to the observer to change continuously (known as the Light Time Effect, LiTE). Nevertheless, studies of its transit timing variations (TTVs) \citep{seeliger2015, maciejewski2018} have been inconclusive so far for the existence of such a potential third body. Only a few studies have attempted to observe the target with high photometric precision to derive its parameters as well as its transit mid-times since its discovery back in 2011 \citep{hartman2011, seeliger2015, maciejewski2018}. However, the system has been observed many times by amateur observers, as a result of which, at least a dozen moderate-quality light curves have been accumulated in the public archive of the Exopolanet Transit Database (http://var2.astro.cz/ETD/), having the potential to be used in a TTV analysis.

We observed the target several times with the 1 meter Turkish telescope T100, located in the Bak{\i}rl{\i}tepe campus of the T\"UB\.{I}TAK National Observatory of Turkey (TUG) at an altitude of 2500 m above sea-level, near the south coast of the country; and achieved very high photometric precision thanks to the well-established telescope defocusing technique \citep{southworth2009, basturk2015}. We have analyzed our own light curves with the state-of-the-art second version of the EXOFAST software paackage \citep{eastman2017, eastman2019}, derived the global parameters of the system making use of the orbital and radial velocity parameters derived by \citet{hartman2011}, and the atmospheric properties of the host star that we have obtained from our own analysis of the Keck / HIRES archival spectra with primarily the iSpec software package \citep{cuaresma2014}. We made use of the brightness of the target in different broadband filters and fit its spectral energy distribution based on its precise distance value thanks to Gaia mission to constrain the radius of its host star  \citep{gaia2016, gaia2018}. Finally, we collected all the light curves obtained so far by a number of observers around the world, all of which we corrected for the barycenter of our Solar System, measured mid-transit times in Dynamical Barycentric Julian Days (BJD-TDB), updated the ephemeris information for the system, established and analyzed the transit timing variation (TTV) diagram. We present the photometric and spectroscopic data that we used and the details of the data reduction procedure in Section \ref{sec:observations}, we provide the information on data analysis and present the parameters of the system as a result of the global modeling of the data, and the transit timing variations in Section \ref{sec:analysis}, and finally discuss the importance of our findings in the context of the hot-Saturn type planets, their formation, orbital evolution, and inflation of their atmospheres, as well as the potential reasons behind the observed transit timing variations in Section \ref{sec:conclusions}.  

\section{Observations and Data Reduction}
\label{sec:observations} 
\subsection{Photometric Data}
We observed seven transits of HAT-P-19b between July 2014 and December 2016 with the 1 m Turkish Telescope T100 and the high quality, cryo-cooler SI 1100 CCD with 4096x4096 pixels attached on it, which has a field of view of $20'$ x $20'$. All the observations have been obtained in the Bessel-$R$ passband with 120 second integration time, which gave the best Signal-to-Noise Ratio (SNR) within an optimal number of observational points for each transit. We only changed the defocusing amount of the telescope, hence the width of the Point Spread Function (PSF) from one night to another according to the sky quality. We observed the same target with higher exposure time (185 seconds) in Bessel-$R$ very recently by defocusing T100 more aggressively to achieve even better SNR. However, thin clouds appeared at high altitudes right after the ingress, made the data acquired at that time totally unreliable causing a gap in a very precise light curve otherwise. We haven't used this light curve neither in the modeling nor in the analysis for its transit timing variations. We provide a log of our observations in Table-\ref{tab:t100_observations}. $\sigma_{ph}$ in the fourth column of this table is the nightly average of the photometric measurement uncertainty of each point on the light curve in milimagnitudes, while  $\sigma_{RMS}$ is the root mean square errors of the linear fit to the out-of-transit segments to detrend the light curves from changing airmass.

Photometric Noise Rate (PNR) \citep{fulton2011} is another indicator of light curve quality defined as the ratio of the standard deviation of the residuals to the median number exposures per minute including also the time spent for read-out. We also provide the $\beta$ factor \citep{winn2008}, that quantifies the correlated noise, and it is defined as the ratio of the average residuals in several bins to the standard deviation of the binned residuals in Table-\ref{tab:t100_observations}. If only the white noise dominates the noise budget, then $\beta$ = 1. The ingress / egress timescale ($\tau$) of HAT-P-19b transits is $\sim$23 minutes. Therefore, we grouped our light curve data points in variable sizes of bins from 13 to 33 minutes and took the median of $\beta$ factors of those 15 bins in total. All our light curves are in the same photometric band (Bessel-R), recorded with the same telescope (T100) with the same exposure time (120 seconds) except our latest observation, when we experimented with the limits of defocusing by making use of 185 second-exposures. Hence, the $\beta$ factors are comparable within the group \citep{winn2008} of our observation runs with T100.  

In all of our observations, $\beta$ factor, which is a good indicator of the red noise, is between 0.85 and 1.33, showing that white noise is the dominant noise source in our observations. We had the largest value for this parameter on observations numbered 4 and 5, with 1.33 and 1.31, respectively. The inferior quality of these light curves is also evident from other statistics, more strikingly from PNR. Therefore, we haven't used these light curves in global modeling. While the $\beta$ factor for the observation numbered 6 is 0.85, when the ingress is missing completely, which complicates the modeling as well as decreases the correlated noise artificially as pointed out by \citet{winn2008}. We had to ignore some images since the linearity limits of the CCD have been exceeded during this observation run as well. Our photometric observations have been affected by the suboptimal weather conditions, all throughout the night on 2015-10-28 and 2016-12-18 (numbered 5 and 7, respectively), increasing the photometric uncertainty to 3-6 milimagnitudes at some point. On 2015-11-05 (number 6), we missed the ingress totally due to bad weather conditions at the time, which made the normalization level and hence the light curve depth somewhat questionable. We also had to ignore the data acquired after the ingress in 2019-08-03 (numbered 8) transit due to the variations in atmospheric conditions. Since these light curves have low-to-moderate quality, either the measured transit depths are significantly different from that observed in the other three nights, when the observing conditions were better or there is an insufficient number of data points at important orbital phases for modeling. Therefore, we decided not to include them in our analysis in obtaining the system parameters. However, it is still possible to determine the mid-transit times with good precision (on the order of a few seconds), hence we used this set of slightly inferior photometric quality observations in the analysis of transit timings except the latest T100 light curve on 2019-08-03. Since the PNR value is a model-independent, hence a convenient measure of white noise, making it possible to compare light curves obtained in different nights/seasons, we used it as a light curve selection criterion for the global modeling. In the first three nights (observation number 1,2,3) we achieved very high precision with the nightly errors around 1 mmag in the Bessel-$R$ filter, and Photon Noise Rate (PNR) values below 0.75. Therefore we selected them to be used in global modeling as a result. Timestamps in T100 FITS files are updated by the computer clock, which is synchronized with a GPS every few seconds. Therefore the errors in the measurements of mid-transit times are dominated by photometric precision rather than the timing.

\begin{table*}
	\centering
	\caption{A log of our photometric observations with T100.}
	\label{tab:t100_observations}
	\begin{tabular}{cccccccc} 
		\hline
		Obs. Number & Starting & Exp.Time & Filter & $\sigma_{ph.}$ & $\sigma_{RMS}$ & PNR & $\beta$\\
                & Date [UT] & [s] &  & [mmag] & [mmag] & [mmag] &  \\ 
		\hline
		1 & 2014-07-28 & 120 & Bessel-$R$ & 1.03 & 0.35 & 0.75 & 1.30 \\
		2 & 2014-08-21 & 120 & Bessel-$R$ & 0.89 & 0.38 & 0.59 & 1.15 \\
		3 & 2015-09-14 & 120 & Bessel-$R$ & 1.10 & 0.47 & 0.71 & 0.96 \\
                4 & 2015-10-04 & 120 & Bessel-$R$ & 1.20 & 0.44 & 1.08 & 1.33 \\
                5 & 2015-10-28 & 120 & Bessel-$R$ & 1.71 & 1.02 & 1.24 & 1.31 \\
                6 & 2015-11-05 & 120 & Bessel-$R$ & 1.24 & 0.66 & 0.94 & 0.85 \\
                7 & 2016-12-18 & 120 & Bessel-$R$ & 0.84 & 0.73 & 0.92 & 0.87 \\
                8 & 2019-08-03 & 185 & Bessel-$R$ & 0.75 & 0.35 & 0.68 & 0.85 \\
		\hline
	\end{tabular}
\end{table*}

In order to achieve high SNR in our photometric measurements, we aggressively defocused T100 in our observations with varying amounts according to the weather conditions in a given night. Telescope defocusing is a well established observing method in the observations of bright targets to increase the photometric precision \citep{southworth2009,basturk2015} by increasing the exposure times. In order not to saturate the detector, short integration times are usually employed during the observations of such bright targets ($m < 13^m$ in our case in the observations with T100), which cause photon noise to dominate. The basic principle of the technique is to increase the exposure times by distributing the Point Spread Function (PSF) of star images over many pixels by defocusing the telescope, hence decreasing the number of photons hitting each pixel in the unit time. Photometric errors due to imperfect flat fielding are also mitigated by an order of magnitude since the pixel-to-pixel variations are averaged out when photons are counted from a larger area. PSF changes due to atmospheric scintillation and imperfect tracking are also subtle when a larger number of resolution elements are involved. An additional advantage of the technique is that the observer wastes much less time due to the read-out procedure, which takes 45 seconds with the CCD attached on T100 for a single image. By increasing the exposure time, however, the observer uses a larger fraction of the total observation time to collect photons from the source.

\begin{figure}
\includegraphics[width=\columnwidth]{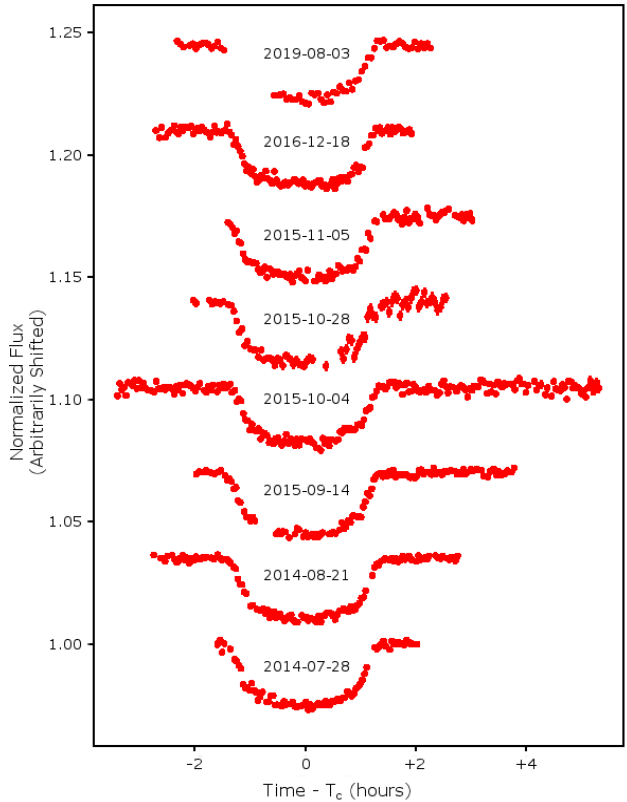}
    \caption{T100 Light Curves of HAT-P-19b Transits in Bessel-$R$ Passband.}
    \label{fig:t100_lc}
\end{figure}

We corrected our images with the AstroImageJ (hereafter AIJ) software package \citep{collins2017} for undesired instrumental effects by making use of the medians of 5-to-10 bias, dark, and twilight-sky flat images, shot during the same night as the observations. We made use of 2x2 pixel-binning to achieve a balance between the photometric precision and timing resolution, which decreases with the increase in the exposure times while the latter gets better since reading over a 4 times smaller number of pixels decreases the read-out time to 15 seconds within our setup. We have converted all the observation timings to Dynamical Barycentric Julian Days (BJD-TDB) and recorded them in the headers of the image files together with the calculated airmass values for those timings. AIJ makes it possible to perform ensemble aperture photometry \citep{honeycutt1992} relative to a number of comparison stars. We used GSC 2283-1197, and 2MASS 00382684+3446556 in the same field with HAT-P-19, as our comparisons in the differential photometry, since their brightnesses are comparable and no photometric variations have been recorded so far in the timescales of our observations and sensitivity limits of our setup. Since our observations are defocused, we determined the center of the apertures ourselves to avoid incorrect positions determined by the centroid method. We employed different aperture sizes for different nights with changing atmospheric seeing values. We then corrected for the airmass, and then normalized the relative fluxes determined by AIJ, by dividing them to the line fit to relative fluxes out of the transit profile. As a result, we obtained the normalized transit light curves for each of the nights of our observations with T100, which we present in the Fig.\ref{fig:t100_lc}.

\subsection{Spectroscopic Data}
High resolution spectroscopic observations of HAT-P-19 have been carried out with High Resolution Ech\'elle Spectrometer (HIRES) attached on the 10 m Keck Telescope between October 2009 and March 2010 by \citet{hartman2011} and the High-Dispersion Spectrograph (HDS) on the Subaru telescope on Mauna Kea to measure its radial velocity variations due to the transiting object, the planetary nature of which has been revealed as a result. For our spectral analysis, we used the only available spectrum in the Keck / HIRES archive, that is free of the absorption lines of $I_2$ gas due to the iodine cell used to increase radial velocity precision because HDS spectra from Subaru Telescope have narrow wavelength coverage and lower SNR. Average spectral resolution of the Keck / HIRES spectrum is R $\sim$ 55000, which is sufficient to determine the fundamental atmospheric properties of the host star. The spectrum covers a wide wavelength interval between 3360 - 8100 $\AA$. An average SNR of $\sim$130 had been achieved in the 1390 second-exposure on 23 February 2010. This spectrum has also been used by \citet{hartman2011} for the same purpose. We corrected the spectrum for both the Doppler shift due to the orbital motion of the star about the common center of mass with the planet HAT-P-19b and the orbital motion of the Earth about the Sun. We have also removed the cosmic rays, and cut the wavelength regions dominated by tellurics and that where the SNR is too low for a thorough analysis. The spectrum has then been normalized to the continuum level, which is determined from a synthetic spectrum used as a visual template for comparison because it is very challenging to determine it from the observational spectrum due to the numerous spectral lines which are blended in most cases. This synthetic spectrum has been designed in iSpec software package \citep{cuaresma2014, cuaresma2019} making use of the lines from Vienna Atomic Line Database (VALD, \citet{piskunov1995, kupka2000, rybichikova2015}) scaled by the ATLAS9 model atmosphere \citep{castellikurucz2004}, and solar chemical abundances from \citet{asplund2009}, and used as a visual aid to determine the continuum. For the computation of this synthetic spectrum Synthe code \citep{kurucz1993}, which is embedded in the iSpec software package, has been used, and the initial values of spectroscopic parameters were taken from \citet{hartman2011}. In the wavelength regions, where there are significant differences between the line strengths in the observed stellar spectrum and the synthetic spectrum, the extent of such a strong line has been accounted for in the determination of the continuum. The continuum points have been marked on the observed stellar spectrum and used as the nodes of cubic spline functions fit to them, and then the spectrum is normalized by dividing the entire spectrum with these spline functions.

\section{Data Analysis and Results}
\label{sec:analysis} 
\subsubsection{Spectrosocpy}
\label{sec:spectroscopy}
\begin{figure}
\includegraphics[width=\columnwidth]{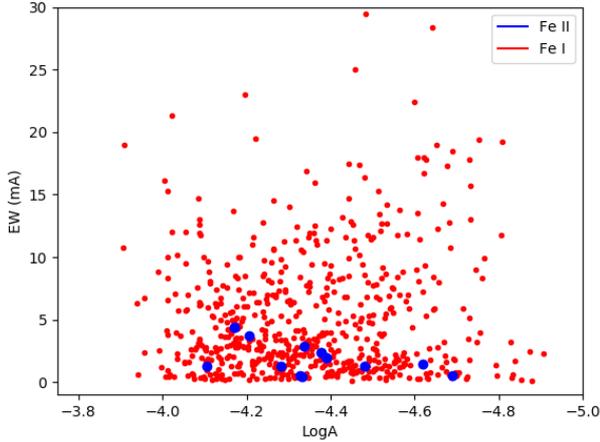}
    \caption{Equivalent widths of 611 Fe-I and 15 Fe-II lines with respect to their abundances for the microturbulent velocity (($v_{mic} = 0.80$ km/s) that minimizes the correlation between the two.}
    \label{fig:curve_of_growth}
\end{figure}

For stars with effective temperatures lower than $5500$ K, techniques based on the measurements of the ratios of equivalent widths or depths of certain lines (mostly Fe-I and Fe-II) \citep{gray1994}, wings of the $H_{\alpha}$ line \citep{fuhrmann2004}, and excitation/ionization balance \citep{santos2004} are not optimum if not totally inadequate in the determination of fundamental parameters due to the excessive line blending \citep{tsantaki2013}. Synthetic spectrum fitting \citep{valenti2005} is the preferred technique in this regime of stars; which has been our main method in deriving the fundamental parameters. We have looked for the best fitting synthetic spectrum to our observational one with the least-squares minimization method by using the iSpec software package \citep{cuaresma2019}. We have tried to fit only the lines of Fe-I and Fe-II having 0.03 to 1.0 line depths, resulted in $\sim$1300 lines which we have further reduced to 626 in total (611 Fe-I and 15 Fe-II) after having eliminated lines with higher uncertainties than EW itself and beyond the observed $\pm$0.5 dex scatter in Fig.\ref{fig:curve_of_growth} about the mean iron abundance. iSpec also filters out the lines from its analysis, to which it fails to fit a Gaussian. Nevertheless, we still end up with somewhat blended lines that we have ignored because it has been shown by \citet{tsantaki2013} that fundamental parameters ($T_{eff}, log~g, [Fe / H]$) derived from an analysis based on only unblended lines to that involving the blended lines significantly differ if equivalent widths of iron lines are employed, especially for stars with effective temperatures lower than $5000$ K. Then resolving the lines and determining the continuum level, and hence measuring the equivalent widths become even a more critical problem. In contrast, we have determined the continuum level by comparing the observed spectrum visually  to a synthetic template, and derived the stellar parameters unambiguously by fitting the best synthetic spectrum.

For an initial fit to the observed data, $T_{eff}$, $log~g$, $[Fe / H]$, and line broadening (the sum of rotational and macroturbulent velocities) have been adjusted. When a solution is converged,   $T_{eff}$, $[Fe / H]$, and broadening amount (in velocity units) have been fixed to the values found in the initial fit, while the surface gravity (log g) has been adjusted to look for a convergence. Since the effect of log g on the spectra of cool dwarfs is marginal, it will be more adequate to determine its value by using the stellar density derived from the light curve analysis. In order to keep the surface gravity consistent with the mean density of the star, that we fixed to its value we derived from the light curve analysis, for each output surface gravity value, we provided all the fundamental parameters to the equation given by \citet{torres2010} as input to derive the stellar mass and radius, hence a mean density value, which we compared to the empirical value that we get from the light curve analysis. When the difference between the two densities is minimized to  $0.001 g / cm^3$, we accepted the current surface gravity as that of the host star and stopped the iteration, while other parameters have been adjusted for a final convergence. The resultant atmospheric model gave the effective temperature ($T_{eff}$), metallicity ($[Fe/H]$), and the broadenings (micro / macroturbulence, and the projected rotational velocity), which are summarized in the third column of Table-\ref{tab:atmospheric_parameters}.

We measured equivalent widths of 611 Fe-I and 15 Fe-II ``clean lines'', and determined the microturbulent velocity ($v_{mic}$), for which the equivalent widths of these lines become independent of the chemical abundances of the related species, and the scatter in the mean abundances for those is minimized. The value we find for this parameter in this manner is not too different for the microturbulent velocities assumed for HAT-P-19 ($v_{mic} = 0.85$ km/s) in the literature based on the calibrations \citep{hartman2011, brewer2016}. However, we preferred to fix the microturbulent velocity to the value we have determined from the curve of growth ($v_{mic} = 0.80$ km/s) given in Fig.\ref{fig:curve_of_growth}.

We have also attempted at deriving the effective temperature from the ionization balance between Fe-I and Fe-II. However, due to the low temperature of the target, and because there are not many lines of Fe-II in the spectrum with good SNR allowing precise measurements to be made, the effective temperature has been underestimated by $\sim$200 K. Hence, we have renounced the analysis and decided to use the fundamental parameters from synthetic spectrum. This kind of analysis based on equivalent width measurements gives reliable microturbulent velocities. That is why we used the $v_{mic}$ we determined from this analysis but not the other parameters.

\begin{figure}
\includegraphics[width=\columnwidth]{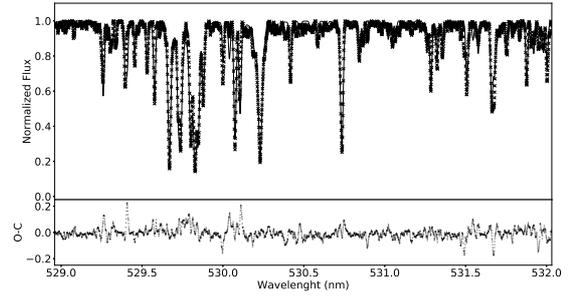}
    \caption{Keck / HIRES spectrum of HAT-P-19 (+) and the best fit synthetic spectrum (-) with iSpec (5290 - 5320 $\AA$ region. A colored version is provided only in the online version.}
    \label{fig:spectrum_best_fit1}
\end{figure}

\begin{figure}
\includegraphics[width=\columnwidth]{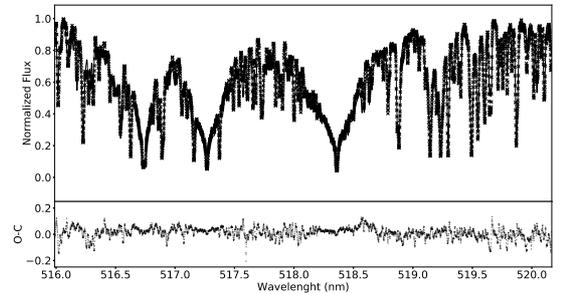}
    \caption{Keck / HIRES spectrum of HAT-P-19 (+) and the best fit synthetic spectrum (-) with iSpec (5160 - 5200 $\AA$ region.}
    \label{fig:spectrum_best_fit2}
\end{figure}

On the other hand, macroturbulent motions and stellar rotation similarly broaden the line profiles, shaping the profiles to be Gaussians. Therefore, the broadening determined by the iSpec software package from the line profiles is the convolution of both broadening mechanisms. If one of these parameters (e.g. mactroturbulence) is fixed to 0, and the other is set free; then the latter (projected rotational velocity ($v~sini$) in our example) is responsible for the entire broadening. That is why we intended to use the typical macroturbulent velocity for our target. Since HAT-P-19 has an effective temperature that is very close to the borderline of the calibration given by \citet{gilmore2012}, we derived the macroturbulent velocity with both calibration equations, one of which is for stars above $T_{eff} = 5000$ K ($v_{mac} = 3.35$ km/s)  and the other below that limit ($v_{mac} = 2.87$ km/s). We found an intermediate value, which is in agreement with the value we found from the weakest Fe-I lines that form deeper in the atmosphere ($v_{mac} = 3.00$ km/s), after having deconvolved the microturbulent velocity from the profiles. We then fixed the macroturbulent velocity to this value, and adjusted the projected rotational velocity and found it to be $vsini = 0.88\pm1.09$ km/s. The derived atmospheric parameters from our analysis of the Keck / HIRES spectra of HAT-P-19 are given in Table-\ref{tab:atmospheric_parameters}, in comparison with the results from previous works. We provide the best synthetic spectrum for the observed in two different wavelength regions in Figs. \ref{fig:spectrum_best_fit1} \& \ref{fig:spectrum_best_fit2}.

\begin{table}
\caption{Atmospheric Parameters of HAT-P-19.}
\label{tab:atmospheric_parameters}
\begin{tabular}{lcccc}  
\hline
\hline
Parameter & H11 & B16 & Initial Fit & Final Fit \\
\hline
$T_{eff}$ [K] & $4990\pm130$ & $4951$ & $4988\pm54$ & $4991\pm50$ \\
$log~g$ [cgs] & $4.54\pm0.05$ & $4.44$ & $4.55\pm0.07$ & $4.53$ \\
$[Fe / H]$ [dex] & $0.23\pm0.08$ & $0.29$ & $0.23\pm0.05$ & $0.24\pm0.05$ \\
$v~sini$ [km/s] & $0.7\pm0.5$ & $1.8$ & $1.41\pm0.87$ & $0.88\pm1.09$ \\
$v_{mac}$ [km/s] & $2.81$ & $1.8$ & $2.82\pm0.59$ & $3.00$ \\
$v_{mic}$ [km/s] & $0.85$ & $0.85$ & $0.79\pm0.15$ & $0.80$ \\
\hline
\end{tabular}
\end{table}

\subsection {Global Modeling}
\label{sec:globalmodel}
We used the state-of-the-art, second version of EXOFAST \citep{eastman2017, eastman2019} in order to obtain a global model of the light curve data, radial velocity parameters (semi-amplitude, eccentricity, and the argument of periastron), stellar atmospheric parameters derived from our spectroscopic analysis, and the brightness of the host star in different passbands. We have fitted our three best light curves from T00, that we selected due to their high quality, quantified by the Photon Noise Rate (PNR) and the $\beta$ parameters as well as their completeness. We enforced Gaussian priors on the epoch ($T_{c} = 2456867.425742$ in BJD-TDB) based on the mid-transit time of our first observation with T100, and orbital period ($P = 4.0087826 $ days) determined by \citet{hartman2011}. We have also made use of Gaussian priors for the radial velocity parameters, semi-amplitude, eccentricity and the argument of periastron centered at the values obtained by \citet{hartman2011} with standard deviations equal to that given in the same study. We used the stellar atmospheric parameters we determined as a result of our own analysis of one high resolution, high SNR Keck / HIRES spectrum, as $T_{eff} = 4991\pm50 K$, $[Fe / H] = 0.24\pm0.05$, and $vsini = 0.88\pm1.09 km/s$ for the mean and standard deviation of Gaussian priors. We haven't enforced a prior for the surface gravity since the light curve modeling of exoplanet transits provide better constraints for this parameter, which is not well constrained in spectroscopic analyses of cool star atmospheres. Limb darkening coefficients for the quadratic law have been interpolated from the tables provided by \citet{claretbloemen2011} during the light curve fitting. Cosine of the orbital inclination (cos i) has been assigned to a uniform prior.

EXOFAST-v2 also models the Spectral Energy Distribution (SED) based on the passband brightnesses from broadband photometry and the stellar distance. We collected passband brightness values of HAT-P-19 from all-sky survey catalogs, which we list in Table-\ref{tab:passband_brightness}. We intended to use the Sloan $ugriz$ magnitudes from the Sloan Digital Sky Survey (SDSS) catalogue. However, there are two different point sources listed in the catalog for the coordinates of our target, while Gaia observations do not reveal a nearby companion to HAT-P-19. This might have been caused by two differing measurements at different times with SDSS. The source closer to the coordinates of HAT-P-19 has conflicting measurements, while that for the source farther away are consistent with its stellar nature. Therefore, we had to use $g'$, $r'$, $i'$ magnitudes from APASS DR9 catalog \citep{henden2016}, while we made use of the $zPS$ from PAN-STARRS catalog \citep{chambers2016} instead, because at least the upper limits for realistic errors have been listed in these catalogs for the passband brightnesses. Finally, we used $u'$ magnitude from the original SDSS catalog \citep{alam2015}, since this value is the same for both point sources in the catalog. We have taken the effective wavelengths of the filters used in the corresponding sky surveys from The Spanish Virtual Observatory (SVO) Filter Service \citep{rodrigo2012,rodrigo2013}, and provided them in Table~\ref{tab:passband_brightness}.

\begin{figure}
\includegraphics[width=\columnwidth]{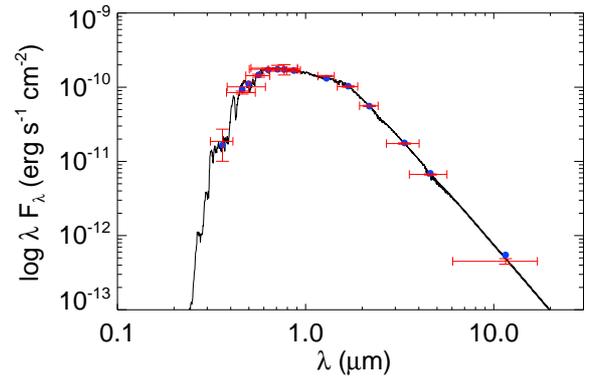}
    \caption{The passband brightnesses are represented with red data points, the error bars of which in wavelength show the widths of the corresponding filters while those in flux denote the measurement uncertainties. The black continuous curve is the model SED, while blue circles are broadband averages on the model.}
    \label{fig:sed_best_fit}
\end{figure}

Our analysis relies on the most precise distance of the object thanks to the exquisite parallax value provided by the Gaia mission \citep{gaia2016, gaia2018}. We accounted for the systematic offset in Gaia parallaxes, noticed by \citet{stassuntorres2018}, in deriving the distance (d = 199.921 $\pm$ 2.738 pc) of HAT-P-19 and added 0.082 miliarcseconds to the Gaia parallax value, and 0.033 to its uncertainty; which translates into a -3.332 pc difference in the distance of the object and +1.271 pc in its uncertainty. EXOFAST constrains the V-band extinction ($A_V$) based on its SED model (Fig.\ref{fig:sed_best_fit} for HAT-P-19), which is a measure of the star's bolometric flux and the ratio of the stellar radius to its distance. Having a precise and independent measurement of the distance, therefore, makes the constraint on the stellar radius very strong through luminosity relation. During the global modeling, the extinction coefficient ($A_V$) has been adjusted but limited by the maximum extinction value to that given by the maps from \citet{schlegel1998}.

\begin{table}
\centering
\caption{Passband Brightnesses of HAT-P-19.}
\label{tab:passband_brightness}
\begin{tabular}{ccc}  
\hline
\hline
Passsband & $\lambda_{eff} \AA$ & Magnitude \\
\hline
\multicolumn{3}{l}{SDSS \citep{alam2015}}\\
\hline
SDSS u' & 3594.9 & $15.589\pm0.100$ \\
\hline
\multicolumn{3}{l}{APASS-DR9 \citep{henden2016}}\\
\hline
Johnson B & 4378.1 & $14.834\pm0.051$ \\
Johnson V & 5466.1 & $12.853\pm0.050$ \\
SDSS g' & 4640.4 & $13.381\pm0.176$ \\
SDSS r' & 6122.3 & $12.500\pm0.043$ \\
SDSS i' & 7439.5 & $12.275\pm0.174$ \\
\hline
\multicolumn{3}{l}{Pan-STARRS \citep{chambers2016}}\\
\hline
zPS & 8657.8 & $12.18\pm0.05$ \\
\hline
\multicolumn{3}{l}{2MASS \citep{cutri2003}}\\
\hline
$J_{2MASS}$ & 12350.0 & $11.095\pm0.020$ \\
$H_{2MASS}$ & 16620.0 & $10.644\pm0.022$ \\
$K_{2MASS}$ & 21590.0 & $10.546\pm0.019$ \\
\hline
\multicolumn{3}{l}{All WISE \citep{cutri2014}}\\
\hline
WISE1 & 33526.0 & $10.495\pm0.022$ \\
WISE2 & 46028.0 & $10.557\pm0.020$ \\
WISE3 & 115608.0 & $10.561\pm0.091$ \\
\hline
\multicolumn{3}{l}{NOMAD \citep{zacharias2005}}\\
\hline
Johnson R & 6695.6 & $11.99\pm0.1$ \\
\hline
\end{tabular}
\end{table}

\begin{figure}
\includegraphics[width=\columnwidth]{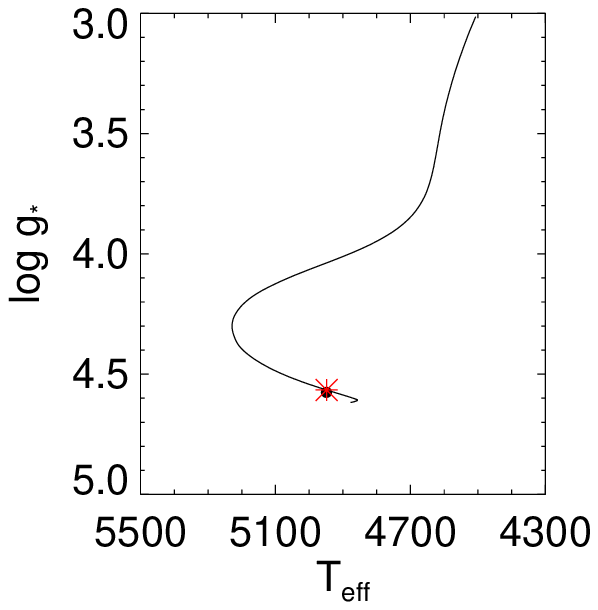}
    \caption{The best fitting MIST track, interpolated at the model values for $M_\ast$, $[Fe/H]_0$ and $EEP$, is given by the black continuous curve. Red asterisk symbol indicates the position along the best fitting MIST track for HAT-P-19, while the black circle is at the model value for $T_{eff}$ and $log~g_\ast$.}
    \label{fig:mist_best_fit}
\end{figure}

EXOFAST-v2 can fit a MESA isochrone (MIST) \citep{paxton2011, paxton2013, paxton2015} and find the position of the host star on the stellar evolutionary tracks based on a trilinear interpolation in Equivalent Evolutionary Phase parameter (EEP), which quantifies the phase at which the host star in its evolutionary history \citep{eastman2019}, initial stellar mass ($M_\ast$) and initial metallicty ($[Fe/H]_0$) parameters, which are the mass and surface metallicity value at the zero-age main-sequence (EEP = 202). From the corresponding MESA stellar track, the code derives the effective temperature ($T_{eff}$), stellar radius ($R_\ast$), and surface metallicity ($[Fe / H]$); and compares them to the model values at the current MCMC step. Since we have a well-constrained $T_{eff}$ - $[Fe / H]$ pair from spectroscopy, and $R_\ast$ from SED fitting, the MCMC algorithm constrains the position of the host star on $T_{eff}$ - $log~g$ plane by changing $M_\ast$, and $[Fe/H]_0$  to match the values of these parameters by keeping $M_\ast$ consistent with that found from the stellar density ($\rho_\ast$). The best-fit MIST evolutionary track for HAT-P-19 from our analysis is given in Fig.\ref{fig:mist_best_fit} with the black continuous curve. 

\begin{figure}
\includegraphics[width=\columnwidth]{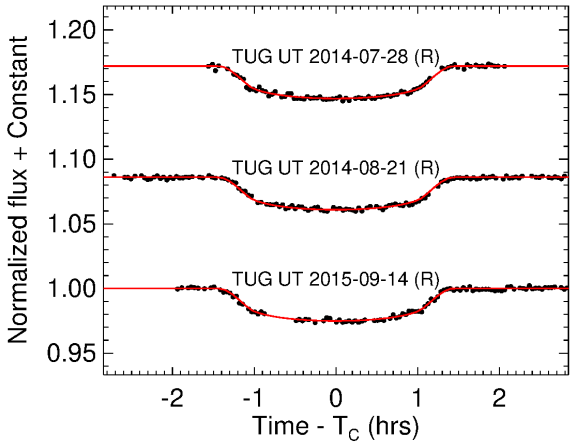}
    \caption{Three selected individual T100 transit observations and the EXOFAST-v2 models to each of these light curves.}
    \label{fig:t100_separate_model}
\end{figure}

\begin{figure}
\includegraphics[width=\columnwidth]{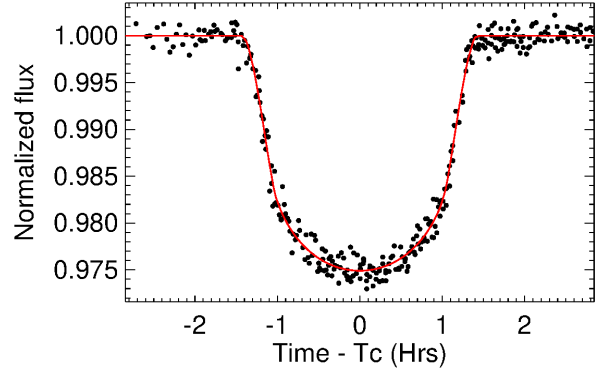}
    \caption{Ensemble light curve of three selected T100 transit observations and the light curve model based on the median parameters given in Table~\ref{tab:global_parameters}.}
    \label{fig:t100_global_model}
\end{figure}

Then the posterior distributions of the global model parameters have been calculated by employing likelihood functions utilizing the goodness of fit estimators ($\chi^2$) from a run of 200 Monte Carlo Markov Chains of 50000 iterations. When the variance between the iterations has become smaller than the variation inherent to the parameter value, the iterations have been aborted by the program, making use of the Gelman-Rubin statistics for the purpose. We provide the median values of all parameters, distributions of which have been determined from the global modeling of three T100 light curves, radial velocity parameters, available passband brightnesses of the target from broadband photometry, Gaia distance corrected for the systematic shift, and stellar atmospheric parameters determined from our own spectroscopic analysis in Table~\ref{tab:global_parameters}. All the light curves used in the global modeling are given separately in Fig.\ref{fig:t100_separate_model}, and in ensemble in Fig.\ref{fig:t100_global_model}.

\providecommand{\bjdtdb}{\ensuremath{\rm {BJD-TDB}}}
\providecommand{\feh}{\ensuremath{\left[{\rm Fe}/{\rm H}\right]}}
\providecommand{\teff}{\ensuremath{T_{\rm eff}}}
\providecommand{\ecosw}{\ensuremath{e\cos{\omega_*}}}
\providecommand{\esinw}{\ensuremath{e\sin{\omega_*}}}
\providecommand{\msun}{\ensuremath{\,M_\odot}}
\providecommand{\rsun}{\ensuremath{\,R_\odot}}
\providecommand{\lsun}{\ensuremath{\,L_\odot}}
\providecommand{\mj}{\ensuremath{\,M_{\rm J}}}
\providecommand{\rj}{\ensuremath{\,R_{\rm J}}}
\providecommand{\me}{\ensuremath{\,M_{\rm E}}}
\providecommand{\re}{\ensuremath{\,R_{\rm E}}}
\providecommand{\fave}{\langle F \rangle}
\providecommand{\fluxcgs}{10$^9$ erg s$^{-1}$ cm$^{-2}$}
\begin{table*}
\label{tab:global_parameters}
\caption{Median values and 1 standard deviations for the parameters of the star HAT-P-19, and its exoplanet HAT-P-19b. \textbf{to be updated by Selcuk with the latest results}}
\setlength{\tabcolsep}{24pt}
\setlength{\extrarowheight}{2.2pt}
\begin{tabular}{llc}
\hline \hline
Symbol & Parameter (Unit) & Value\\
\hline
\multicolumn{3}{l}{Stellar Parameters:} \\
\hline
\\
$M_*$&Mass (\msun)&$0.834^{+0.034}_{-0.031}$\\
$R_*$&Radius (\rsun)&$0.7879^{+0.093}_{-0.091}$\\
$L_*$&Luminosity (\lsun)&$0.399^{+0.013}_{-0.013}$\\
$\rho_*$&Density (cgs)&$2.41^{+0.12}_{-0.11}$\\
$\log{g}$&Surface gravity (cgs)&$4.566\pm0.018$\\
$T_{\rm eff}$&Effective Temperature (K)&$4961^{+40}_{-43}$\\
$[{\rm Fe/H}]$&Metallicity (dex)&$0.247\pm0.05$\\
$[{\rm Fe/H}]_{0}$&Initial Metallicity &$0.227^{+0.055}_{-0.056}$\\
$Age$&Age (Gyr)&$6.2^{+4.7}_{-4.0}$\\
$A_V$&V-band extinction (mag)&$0.170^{+0.054}_{-0.066}$\\
$d$&Distance (pc)&$200.4^{+2.3}_{-2.2}$\\
\\
\hline
\multicolumn{3}{l}{Planetary Parameters:}\\
\hline
\\
$P$&Period (days)&$4.00878236^{+0.00000050}_{-0.00000049}$\\
$R_P$&Radius (\rj)&$1.089^{+0.018}_{-0.018}$\\
$T_C$&Time of conjunction (\bjdtdb)&$2456867.42602\pm0.00015$\\
$T_0$&Optimal conjunction Time (\bjdtdb)&$2457031.78609\pm0.00015$\\
$a$&Semi-major axis (AU)&$0.04649^{+0.00062}_{-0.00059}$\\
$i$&Inclination (Degrees)&$88.67^{+0.41}_{-0.25}$\\
$e$&Eccentricity &$0.084\pm0.041$\\
$\omega_*$&Argument of Periastron (Degrees)&$-90\pm43$\\
$T_{eq}$&Equilibrium temperature (K)&$984\pm10$\\
$M_P$&Mass (\mj)&$0.290\pm0.016$\\
$K$&RV semi-amplitude (m/s)&$42.1^{+2.0}_{-2.0}$\\
$R_P/R_*$&Radius of planet in stellar radii &$0.1421^{+0.0015}_{-0.0017}$\\
$a/R_*$&Semi-major axis in stellar radii &$12.69^{+0.21}_{-0.20}$\\
$\delta$&Transit depth (fraction)&$0.02019^{+0.00043}_{-0.00047}$\\
$\tau$&Ingress/egress transit duration (days)&$0.01613^{+0.00097}_{-0.0010}$\\
$T_{14}$&Total transit duration (days)&$0.11804^{+0.0049}_{-0.0043}$\\
$b$&Transit Impact parameter &$0.313^{+0.065}_{-0.01}$\\
$\tau_S$&Ingress/egress eclipse duration (days)&$0.01391^{+0.00051}_{-0.00050}$\\
$T_{S,14}$&Total eclipse duration (days)&$0.1043^{+0.0049}_{-0.0043}$\\
$\rho_P$&Density (cgs)&$0.279^{+0.021}_{-0.019}$\\
$logg_P$&Surface gravity &$2.783^{+0.027}_{-0.027}$\\
$\Theta$&Safronov Number &$0.0297\pm0.0015$\\
$\fave$&Incident Flux (\fluxcgs)&$0.2114^{+0.0091}_{-0.0089}$\\
$M_P/M_*$&Mass ratio &$0.000332\pm0.000017$\\
$d/R_*$&Separation at mid transit &$13.49^{+0.51}_{-0.50}$\\
$P_T$&A priori non-grazing transit prob &$0.0636^{+0.0025}_{-0.0023}$\\
$P_{T,G}$&A priori transit prob &$0.0847^{+0.0032}_{-0.0030}$\\
$P_S$&A priori non-grazing eclipse prob &$0.0727^{+0.0018}_{-0.0019}$\\
$P_{S,G}$&A priori eclipse prob &$0.0967^{+0.0027}_{-0.0028}$\\
\\
\hline
\multicolumn{3}{l}{Wavelength Parameters (R):}\\
\hline
\\
$u_{1}$&linear limb-darkening coeff &$0.573\pm0.024$\\
$u_{2}$&quadratic limb-darkening coeff &$0.173\pm0.029$\\
\\
\hline
\end{tabular}
\end{table*}

\subsection{Transit Timing Variations}
\label{sec:ttv}
We have collected light curve data for HAT-P-19b transits from the literature and the Exoplanet Transit Database (ETD, http://var2.astro.cz/ETD/), where observations of amateur astronomers are also published. We contacted all the observers with quality transit light curves (indicated with a data quality index 2 out of 5 or better (1) as given in ETD). Based on these communications, we only included light curves from observers, who have made use of either a GPS device or a timing server to coordinate their timings throughout their observations, and who have provided the time reference (geocentric JD-UTC, HJD, BJD-UTC, etc.) they present their observations in. In the absence of such information, we asked for the raw data from the observer, reduced ourselves, computed and compared the light curves and the mid-transit times that we obtained with that given ETD by the observer. We converted the timings of all observations from the timing reference frames in which they were recorded to dynamic barycentric julian days (BJD-TDB) by using our own scripts that we developed based on the relevant modules and functions of the astropy package \citep{astropy2013, astropy2018}. As a result, we had a homogenous set of mid-transit times covering almost 10 years and 889 orbital periods since the discovery of the exoplanet HAT-P-19b.

\begin{table*}
\centering
\caption{Mid-transit times of HAT-P-19b from the literature and Exoplanet Transit Database (ETD).}
\label{tab:ttv_data}
\setlength\tabcolsep{12pt}
\begin{tabular}{ccccc}  
\hline
\hline
Mid-Transit Time & Error & Filter & Observer / & RMS of Transit Fit \\
(BJD-TDB) & (days)  &   & Reference & (Norm. Flux Units) \\
\hline
2455135.630769 & 0.000303 & Sloan $i$ & \citet{hartman2011} & 0.0015 \\
2455167.701479 & 0.000478 & Sloan $i$ & \citet{hartman2011} & 0.0030 \\
2455480.386468 & 0.000903 & Clear & Lomoz F. & 0.0053 \\
2455496.419664 & 0.001095 & $R_c$ & Naves R. & 0.0035 \\
2455524.482556 & 0.000326 & Clear & Muler G. & 0.0020 \\
2455528.492200 & 0.000385 & Clear & Ruiz J. & 0.0027 \\
2455885.275270 & 0.000472 & Clear & Ayiomamitis A. & 0.0034 \\
2455889.283096 & 0.000237 & Bessel R & \cite{seeliger2015} & 0.0019 \\
2455905.317866 & 0.000273 & Bessel R & \cite{seeliger2015} & 0.0021 \\
2455913.336275 & 0.000175 & $R_{c}$ & \cite{seeliger2015} & 0.0012 \\
2455921.350822 & 0.000825 & $R_c$ & Naves R. & 0.0027 \\
2455921.352820 & 0.001175 & Clear & Corfini G. & 0.0031 \\
2455937.384273 & 0.001320 & $R_c$ & Naves R. & 0.0047 \\
2456145.843820 & 0.000534 & Clear & Shadic S. & 0.0037 \\
2456149.853010 & 0.000730 & Clear & Shadic S. & 0.0047 \\
2456173.906160 & 0.000500 & Clear & Garlitz J. & 0.0040 \\
2456270.115655 & 0.000436 & $R_c$ & Liyun Z. & 0.0028 \\
2456494.608824 & 0.001018 & Clear & Gonzalez J. & 0.0049 \\
2456867.425742 & 0.000201 & Bessel R & T100 (this study) & 0.0010 \\
2456887.472620 & 0.001220 & V & Horta F. G. & 0.0046 \\
2456891.479834 & 0.000184 & Bessel R & T100 (this study) & 0.0009 \\
2456899.494510 & 0.000932 & $R_c$ & Barbieri L. & 0.0069 \\
2456927.557710 & 0.000405 & $R_c$ & Gillier C. & 0.0018 \\
2456943.593630 & 0.000429 & Clear & Benni P. & 0.0043 \\
2456935.575712 & 0.000330 & Bessel R & \cite{seeliger2015} & 0.0025 \\
2456947.602450 & 0.000828 & $R_c$ & Horta F. G. & 0.0025 \\
2456951.612756 & 0.001464 & $I_c$ & Shadic S. & 0.0099 \\
2456975.666389 & 0.001464 & $I_c$ & Shadic S. & 0.0099 \\
2457280.330264 & 0.000233 & Bessel R & T100 (this study) & 0.0010 \\
2457284.338780 & 0.000418 & Clear & Ogmen Y. & 0.0030 \\
2457300.375061 & 0.000305 & Clear & \cite{maciejewski2018} & 0.0020 \\
2457300.375138 & 0.000249 & Bessel R & T100 (this study) & 0.0011 \\
2457304.383495 & 0.000259 & Clear & \cite{maciejewski2018} & 0.0019 \\ 
2457316.408880 & 0.000463 & $R_c$ & Salisbury M. & 0.0022 \\
2457324.426892 & 0.000452 & Bessel R & T100 (this study) & 0.0020 \\
2457328.436220 & 0.000330 & Clear & Bretton M. & 0.0016 \\
2457332.444435 & 0.000290 & Bessel R & T100 (this study) & 0.0014 \\
2457336.454720 & 0.000516 & Clear & Bretton M. & 0.0025 \\
2457340.462366 & 0.000655 & Clear & Molina D. & 0.0027 \\
2457356.501420 & 0.000896 & Clear & Bretton M. & 0.0031 \\
2457725.306940 & 0.000192 & Clear & Bretton M. & 0.0013 \\
2457725.308280 & 0.000488 & Clear & Bretton M. & 0.0020 \\
2457729.313670 & 0.000492 & Clear & Bretton M. & 0.0022 \\
2457733.324840 & 0.000522 & Clear & Bretton M. & 0.0023 \\
2457745.351760 & 0.000369 & Clear & Girardin E. & 0.0061 \\
2457753.367410 & 0.000519 & Clear & Bretton M. & 0.0032 \\
2458699.536574 & 0.000430 & Bessel R & T100 (this study) & 0.0012 \\

\hline
\end{tabular}
\end{table*}

\begin{figure}
\includegraphics[width=\columnwidth]{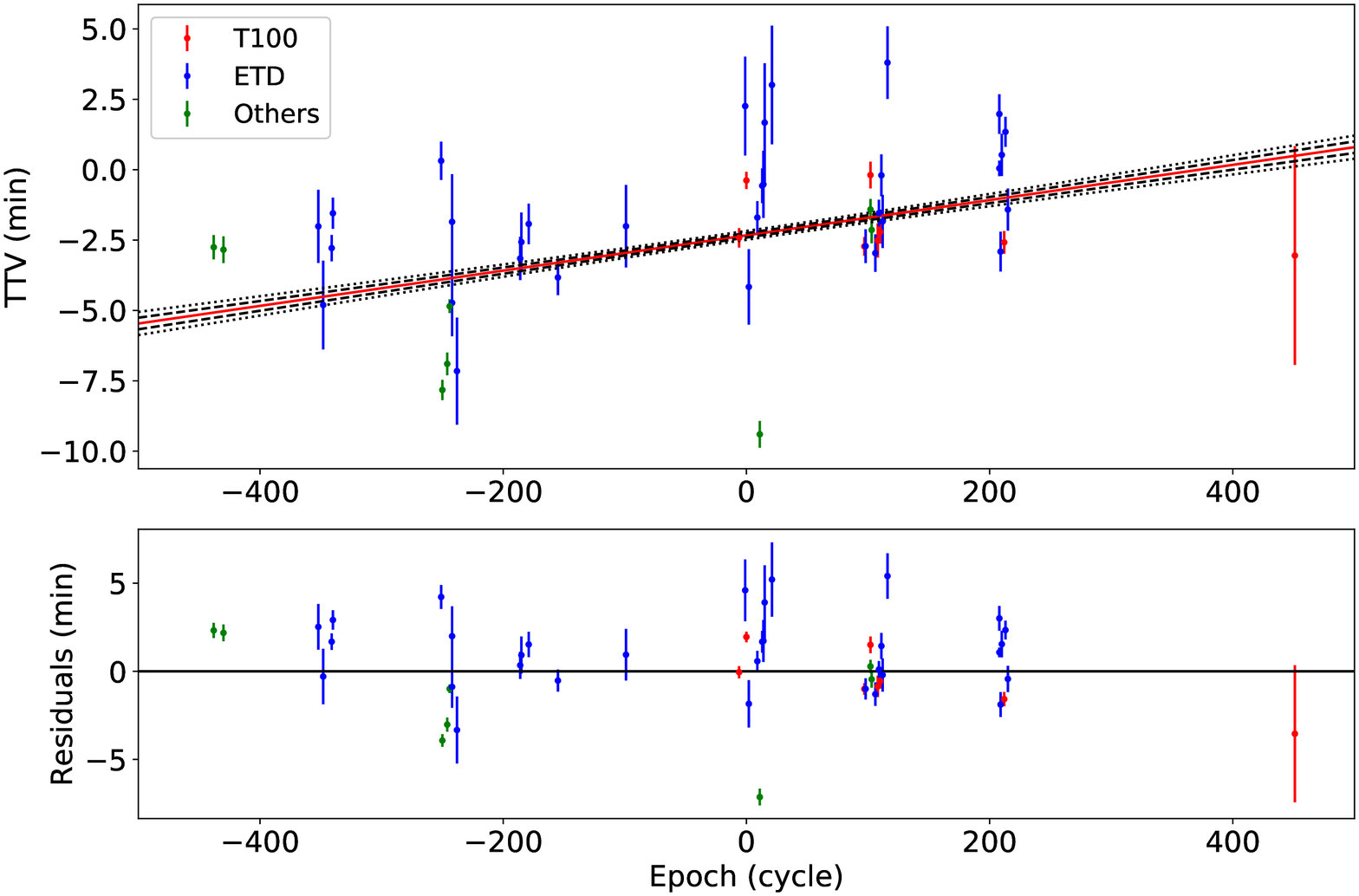}
\caption{Difference between the observed and calculated mid-transit times with respect to that measured from our own observation with T100 (2014-08-21, 2456891.479834(184)) and the orbital period provided by \citet{hartman2011} for HAT-P-19, our linear fit (red continuous) and its 1 (dashed) and 2 (dotted) standard deviations, based on ETD (blue), T100 (red), and other literature data (green).}
    \label{fig:ttv_ETD_fit}
\end{figure}

\begin{figure}
\includegraphics[width=\columnwidth]{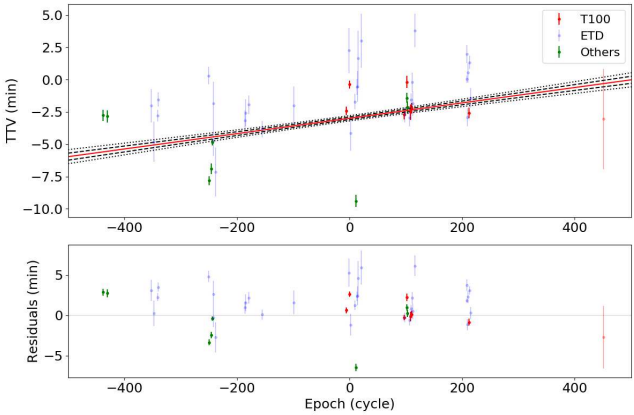}
    \caption{The same as Fig-\ref{fig:ttv_ETD_fit}, but ETD data have been ignored during the fitting, yet plotted for comparison with the Fig-\ref{fig:ttv_ETD_fit}.}
    \label{fig:ttv_noETD_fit}
\end{figure}

We then measured the mid-transit times of all the light curves that we collected (summarized in Table-\ref{tab:ttv_data}) from the literature and ETD, as well as our all eight T100 light curves including the five that we haven't used to determine the parameters due to larger photometric errors and interruptions during the observations. We made use of the first EXOFAST version \citep{eastman2017} for speed to fit all these light curves separately to measure the mid-transit times in BJD-TDB, forming a homogeneous data-set from a heterogeneous set of observations. However, we decided to exclude the latest T100 light curve from the sample (2019-08-03), which gave a mid-transit time with the highest uncertainty due to the large gap in the data making it impossible to determine the timing of the ingress. Then we have taken the differences between observed mid-transit times and expected times of mid-transits based on the mid-transit time we have measured from our best light curve with the minimum scatter (light curve recorded on 2014-08-21 (number 2), see Fig.\ref{fig:t100_lc} and Table-\ref{tab:t100_observations}) and the orbital period given by \citet{hartman2011}. We then plotted these differences with respect to cycle number (epoch) of each observed transit before and after the reference mid-time, and prepared two so-called transit timing variation (TTV) plots, one with a linear fit to all data points including the ETD data, one excluding them (given in Fig.\ref{fig:ttv_ETD_fit}, and Fig.\ref{fig:ttv_noETD_fit}, respectively) Although we ignored the mid-transit time determined from our latest T100 light curve in the analysis, we plotted it on the TTV diagram (last red data point with the largest bars at the end of both TTV diagrams) and listed in Table-\ref{tab:ttv_data} for future reference since it is significantly below the linear trend and it is the only observation made and reported this year. 

\begin{figure}
\includegraphics[width=\columnwidth]{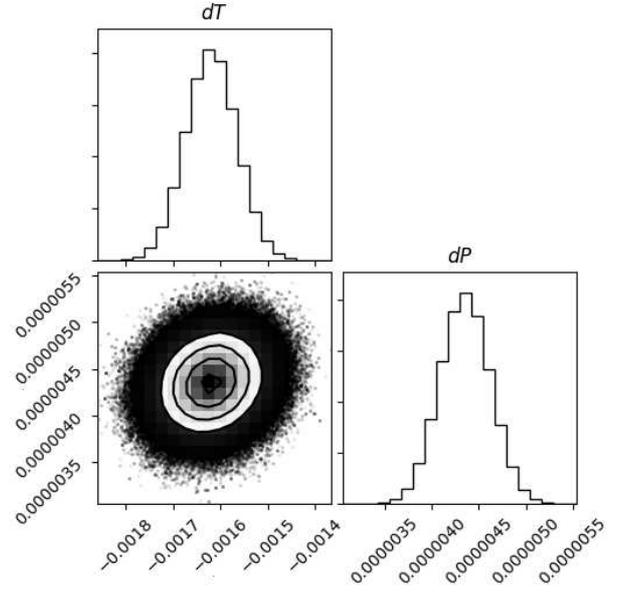}
    \caption{Corner plot showing the covariance between the orbital period and the mid-transit time, and the histograms of probability distribution for both parameters.}
    \label{fig:ttv_corner_plot}
\end{figure}

We took random, equally probable samples from a parameter space of linear coefficients (slope and y-intercept) and computed the likelihood of a linear fit with these coefficients to both TTV diagrams. As a result we have obtained the posterior probabilities for them (Fig.\ref{fig:ttv_corner_plot}, corresponding to the change in reference mid-transit time and the orbital period within an MCMC run, involving 5000 number of iterations and 500 walkers. We have discarded the first 500 steps (the so-called burn-in period) in each of the random walkers until an equilibrium is settled in the search. We decided to use the ETD data in our TTV analysis because there are not many data points left (only 12 including our own) once they are excluded. In addition, the slopes of the linear fits to both data sets (including ETD data and excluding them, there are only 0.015 seconds between the two) are so similar that, the difference amounts to less than a minute (53.20 seconds) after a thousand orbital periods, no matter which one is selected as a reference. Therefore we have refined the ephemeris information from this analysis based on all accepted data from the ETD, literature, and our T100 light curves (Eq.\ref{eq:ephemeris}).

\begin{equation}
\scriptsize{
T_{c}\,(\textnormal{BJD-TDB}) = 2,456,891.478267(55) + 4^{d}.00878862(27) \times E}
\label{eq:ephemeris}
\end{equation}

\begin{figure}
\includegraphics[width=\columnwidth]{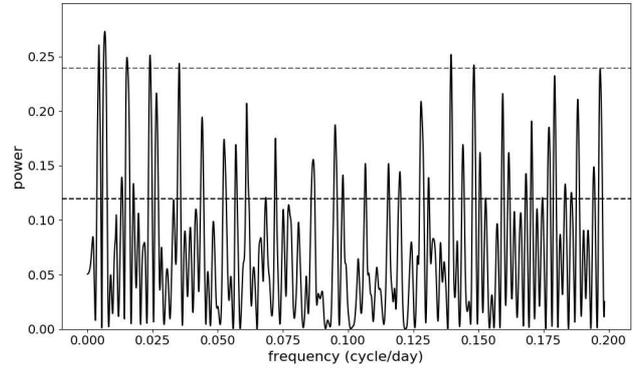}
    \caption{Lomb-Scargle Periodogram of TTV residuals from the linear fit in Fig-\ref{fig:ttv_ETD_fit}. 1 and 2 standard deviations are indicated with dashed and dotted horizontal lines.}
    \label{fig:ttv_LS}
\end{figure}

We have also searched for a periodicity in the frequency space (Fig.\ref{fig:ttv_LS}), and have not found one in 2 standard deviations, the peaks over which are most probably due to the fact that the data is sparse and unevenly distributed in time since the orbital period is almost an integer ($\sim 4.00$ days) making the transits only observable between July and September from the ground since its discovery. The direct consequence of the Fourier analysis is that an unseen third body, gravitationally bound to the system, with an orbital period less than the time span of all observations (7.167 years) and with a mass $m_2~sini > 37$ M$_{jup}$ would be detectable, orbiting the star at $a~sini = 3.40$ AU distance. These parameters set the lower detection limits from the TTV diagrams in Figs. \ref{fig:ttv_ETD_fit} \& \ref{fig:ttv_noETD_fit}. Further observations will be needed to comment on the nature of the observed transit timing variations in the system.

\section{Conclusions and Discussion}
\label{sec:conclusions}
HAT-P-19b has some key characteristics, that make it an important target for follow-up observations and further analysis. First of all, the residuals of its radial velocity data from its orbital solution that follow a linear trend may be an indicator of a gravitationally bound, yet unseen object. Such an object would cause Transit Timing Variations (TTVs) detectable with long-baseline photometry from the ground and space. Non-zero eccentricity ($e = 0.084$) of its short-period orbit supports such an expectation since orbital circularization timescales are significantly shorter compared to the stellar age, especially for Saturn-mass planets such as HAT-P-19b orbiting a $\sim6$ Gyr old star \citep{mills2017, rampalli2019}. However, the orbital eccentricity is only $e = 0.084$ with an uncertainty of half of its derived value. The alarming observation may be the radial velocity residuals being within a level of 60 m/s \citep{hartman2011}. However, we have not found a statistically significant ($> 2\sigma$) periodic signal in our Fourier analysis after having removed the linear trend observed in its TTV diagram due to the accumulation of uncertainties in the reference elements ($T_0$ and $P_{orb}$). Nevertheless, we suggest amateur and professional observers continue their observations of its transits since a longer baseline variation is possible. Because the observed amplitude on the O-C diagram is larger than twice the standard deviation of observations, the TTV signal seems to be real, and hence, should have a physical reason. A very good candidate for such a reason is the activity of the star, which was reported to cause a wavelength-dependent photometric variation of 3.0 to 4.7 milimagnitudes from I to V band, periodically within $35.5 \pm 2.5$ days \citep{mallonn2015}. We haven't detected such a periodicity in our Fourier analysis within 2 $\sigma$. However, the observed TTVs might have been caused by either the imperfect measurements of transit timings from asymmetric transit profiles due to stellar variability or imperfect weather conditions at the time, which affect only some of the light curves in our sample obtained with good sensitivity, or quasi-periodic changes in the activity level within a longer cycle. Both aspects of activity phenomenon could also have caused the observed residuals from the Keplerian fit in the radial velocity measurements, the amplitude of which ($\sim 60 m/s$) is also in agreement with an activity level potentially producing the observed amplitude of photometric variations in V and I bands. Although there is no correlation between these measurements and the bisector inverse span (BIS) \citep{hartman2011}, the length of an activity cycle is incomparably longer with respect to the orbital period, and bright regions around colder stellar spots might have an opposite effect on the bisector span, negating the spot induced radial velocity variations but not affecting the brightness with a sufficient amount in the short run.

Within this work, we have analyzed a high-resolution Keck / HIRES spectrum to derive the atmospheric parameters of the host star HAT-P-19. We have used the results as input parameters for the global modeling based on our high-quality transit light curves, radial velocity semi-amplitude and brightness of the star in different passbands that we collected from all-sky survey catalogs to obtain the final parameters of the planet HAT-P-19b. We made use of the latest (second) version of EXOFAST software package \citep{eastman2017, eastman2019} for the global modeling. Since we have an ultra-precise measurement of the stellar parallax, our SED modeling was based on the most precise distance, which enabled us to derive the stellar radius semi-empirically when the small correction applied by fitting an evolutionary track with a MESA isochrone (MIST) is also accounted for. This resulted in a more precise and accurate value for the planetary radius, which depends less on the theory of stellar evolution. Although the distance value derived from the distance modulus based on absolute and apparent magnitudes of the target in K-band ($d = 215 \pm 15 pc$) \citep{hartman2011} is not too different from that derived from Gaia parallax ($d \sim 200.8 ^{+2.7}_{-2.6} ~ pc$), which is part of the reason why our final stellar and planetary parameters are similar to that published in the discovery paper by \citet{hartman2011}, the agreement is barely within $\pm 1 \sigma$ of the uncertainties. 

Based on these parameters and the definitions by \citet{kreidberg2018}, we calculated the scale height to be $H \sim 582$ km and the amplitude of the absorption signal as $422 \pm 18$ ppm at $3.6 \mu m$, and $735 \pm 28$ ppm at $4.5 \mu m$, which makes it a very good candidate for James Webb Space Telescope to constrain its atmospheric properties, for which it is already in the prime target list \citep{moliere2017}.

Stellar parameters of $M$ and $K$ dwarfs are subject to only slight changes during their main sequence lifetimes, which significantly increases the uncertainties on stellar ages derived from isochrone fitting. We found a smaller age value for HAT-P-19 $6.2^{+4.7}_{-4.0} Gyrs$ than found by \citet{hartman2011} ($8.8 \pm 5.2 Gyrs$). The difference is very subtle. However, \citet{mallonn2015} determined a gyrochronological age of   $5.5^{+1.8}_{-1.3}$ Gyrs based on the rotation period of $P_{rot} = 35.5 \sim 2.5$ days derived from the out-of-transit variation of the host star due to star spot-induced brightness changes modulated with the rotation. Our estimate for the stellar age is in agreement with their result from their photometric campaign observations within the limits of their uncertainties.

Hot-Saturn type planets on short-period orbits are also interesting in their own rights, because they are at the limit of a region in which there is a dearth of planets so-called the sub-Jovian desert \citep{mazeh2016, szabo2011}. Planets with parameters close to but above this limit, such as HAT-P-19b, WASP-49b, WASP-147b, and WASP-160Bb are thought to conserve their bloated, outer volatile layers \citep{mordasini2015} while planets below that limit lose their atmospheres and end up at the bottom of the desert as a small naked core \citep{nielsen2019}. This can explain why the exoplanet HAT-P-19b does not follow the planet radius-host star metallicity correlation as given by \citet{enoch2011}, which was noted by \citet{hartman2011} for the particular cases of HAT-P-18b and HAT-P-19b. The planet radius is expected to inversely proportional to host star metallicity, because the metal content of the planet would lead to a larger core and a smaller radius. However, it might have formed beyond the snow-line of the protoplanetary disc with smaller cores where the material is poor in metal content, and then the atmospheres of these planets might have gathered some gas during their migration in the disk. During their travel inwards, they become subject to more and more irradiance as they come closer and closer to their host stars, which bloats their atmospheres due to internal heating \citep{jackson2008}, as a result of which they end up as low-density planets where their orbits become stable. Since the gyrochroological age found for HAT-P-19 ($\sim 5.5$ Gyrs) \citep{mallonn2015} is compatible with what we have found from MESA isochrones ($\sim 6.2$) Gyrs, a disk driven migration is a more plausible migration scenario \citep{lin1996, ward1997} than a high-eccentricity migration \citep{rasio1996, fabrycky2007}, in which angular momentum transfer from the planet to the star would cause an increase in the rotation rate, which we do not observe in HAT-P-19 rotating at the projected rate of $v sini \sim0.88$ km/s \citep{demangeon2018}. The in-situ formation scenario is also found to be plausible for hot-Jupiters and hot-Saturns \citep{bailey2018} lately. Then the kinetic heating \citep{guillot2002} might be the key mechanism for the inflation of their atmospheres. However, explaining why the planet ended up with as small a core as that of HAT-P-19, will be a challenge, considering the observed high-metallicity of the star, expected to enrich its vicinity. All the parameters we have derived within this study points to a disk-driven migration scenario after the formation of the planet beyond the snow-line as the most likely scenario for its orbital, structural, and atmospheric evolution.

\section*{Acknowledgements}

We gratefully acknowledge the support by The Scientific and Technological Research Council of Turkey (T\"UB\.{I}TAK) with the project 116F350. We thank (T\"UB\.{I}TAK) for partial support in using T100 telescope with project numbers 12CT100-378 and 16CT100-1096. This research has made use of the SVO Filter Profile Service (http://svo2.cab.inta-csic.es/theory/fps/) supported by the Spanish MINECO through grant AYA2017-84089. Some of the data presented herein were obtained at the W. M. Keck Observatory, which is operated as a scientific partnership among the California Institute of Technology, the University of California and the National Aeronautics and Space Administration. The Observatory was made possible by the generous financial support of the W. M. Keck Foundation. We thank all the amateur and professional observers who report their data to Exoplanet Transit Database (ETD), permitted us to use them, and answered all of our never-ending questions.












\bsp	
\label{lastpage}
\end{document}